\documentclass[12pt]{iopart}
\usepackage{iopams}
\usepackage{epsf}
\begin{document}

\title [Collisional and thermal ionization of sodium Rydberg
atoms I]{Collisional and thermal ionization of sodium Rydberg
atoms I. \\ Experiment for \textit{n}S and \textit{n}D atoms with
\textit{n}=8-20}

\author{I.~I.~Ryabtsev$^1$\footnote[3]{To whom correspondence should be addressed
(ryabtsev@isp.nsc.ru)}, D.~B.~Tretyakov$^1$,  I.~I.~Beterov$^1$,\\
N.~N.~Bezuglov$^2$, K.~Miculis$^3$, A.~Ekers$^{3,4}$}
\address{$^1$Institute of Semiconductor Physics, Department of Quantum Electronics, Pr.Lavrentyeva 13,
630090 Novosibirsk, Russia }
\address{$^2$St.Petersburg State University, Fock Institute of Physics, 198904 St. Petersburg, Russia}
\address{$^3$University of Latvia, Institute of Atomic Physics and Spectroscopy, LV-1586 Riga, Latvia}
\address{$^4$University of Kaiserslautern, Department of Physics, D-67663 Kaiserslautern, Germany}

\begin{abstract}
Collisional and thermal ionization of sodium \textit{n}S and \textit{n}D
Rydberg atoms with \textit{n}=8-20 has been studied. The experiments were
performed using a two-step pulsed laser excitation in an effusive atomic
beam at atom density of about 2$ \times $10$^{10}$ cm$^{-3}$. Molecular and
atomic ions from associative, Penning, and thermal ionization processes were
detected. It has been found that the atomic ions were created mainly due to
photoionization of Rydberg atoms by photons of blackbody radiation at
the ambient temperature of 300K. Blackbody ionization rates and effective
lifetimes of Rydberg states of interest were determined. The molecular ions
were found to be from associative ionization in
Na(\textit{nL})+Na(3S) collisions. Rate constants of
associative ionization have been measured using an original method based on
relative measurements of Na$_{2}^{+}$ and Na$^{+}$ ion
signals.
\end{abstract}
 \maketitle

\section{Introduction}

Collisional ionization of atoms may often be associated with highly excited quasi-molecular systems, which are
formed temporarily in course of the collision. Such quasi-molecules may, at certain internuclear distances, couple
to the ionization continuum and emit an electron. Depending on the total excitation energy of the system and
relative positioning of the covalent and ionic molecular potential curves, the quasi-molecule may end up as a
stable molecular ion (associative ionization), or ionize and dissociate into an atomic ion and a neutral atom
(Penning process). Conventional theory \cite{Janev1980,Mihajlov1981,DumanShmatov} treats such processes
deterministically in terms of interaction of a few covalent bound states with the ionization continuum. In the
reality, however, the incoming covalent state interacts with a large number of other molecular Rydberg states.
Moreover, since the density of Rydberg states is very high, it may be impossible to single out separate
interactions between two Rydberg states at a time. Obviously, under such conditions it is not practicable to apply
the exact quantum mechanical description. This drawback of the deterministic description of ionization dynamics
can be overcome by involving a new alternative approach, the so-called stochastic ionization model
\cite{Bezuglov1999}.

The stochastic model describes the collisions with Rydberg atoms as a chaotic migration of highly excited electron
accompanying the motion of both colliding nuclei. The electron ``diffuses'' through the dense energy spectrum of
the highly excited quasi-molecule, eventually ending up in the ionization continuum. The stochastic theory has
already yielded encouraging results by properly describing some of the earlier experimental results on associative
ionization involving Rydberg atoms \cite{BezuglovEkers2002,Bezuglov2003,Bezuglov2002}.

The aim of the present work was to test experimentally the ranges of validity of the stochastic theory developed
in \cite{Bezuglov1999,BezuglovEkers2002,Bezuglov2003,Bezuglov2002,Bezuglov1997}. For this purpose, we studied the
ionizing collisions of sodium Rydberg atoms Na(\textit{nL}) with atoms in the 3S ground state Na(3S). In what
follows we describe the measurements of ionization signals occurring after excitation of sodium atoms to
\textit{n}S and \textit{n}D Rydberg states with \textit{n}=8-20 in a single atomic beam at a source temperature of
635 K. To the best of our knowledge, there are no data in the literature on ionization of sodium atoms in
\textit{n}S and \textit{n}D states for this range of principal quantum numbers. In particular, the processes of
interest are:

\begin{center}
\begin{tabular*}{0.7 \textwidth}{@{\extracolsep{\fill}} l l l}
\\
$\mbox{Na}(nL)+\mbox{Na(3S)}\rightarrow \mbox{Na}^+_2+e$ & Associative Ionization (AI); & \\
\\
$\mbox{Na}(nL)+\mbox{Na(3S)}\rightarrow \mbox{Na}^++
\mbox{Na(3S)}+e$&Penning Ionization (PI); & \\
\\
\end{tabular*}
\end{center}

\noindent where Na$^{+}$ is an atomic ion, and Na$_{2}^{+}$ is a molecular ion.

The earlier experiments on collisional ionization of Na Rydberg atoms are reported in
\cite{Boulmer1983,ZagrebinSamsonZ,ZagrebinSamsonJ,Weiner1986,Burkhardt1984,Burkhardt1986,Klucharev}. The
measurements of absolute rate constants have been done in \cite{Boulmer1983} for \textit{n}P states $(n=5-15)$ in
a crossed beam experiment at source temperatures of 600 K and pulsed laser excitation. The rate constants for
\textit{n}P states $(n=5-21)$ in a single atomic beam at 700 K source temperature and continuous excitation were
reported in \cite{ZagrebinSamsonZ,ZagrebinSamsonJ}. The measurements for \textit{n}S, \textit{n}P, and
\textit{nL}($L>1$) states $(n=17-27)$ in a single Na beam at 1000 K and pulsed laser excitation have been made in
\cite{Weiner1986}. The use of pulsed excitation in \cite{Boulmer1983,Weiner1986} allowed the authors to
distinguish between AI and PI channels and to determine the AI rate constants alone, while only the total (AI+PI)
rates could be measured in \cite{ZagrebinSamsonZ,ZagrebinSamsonJ}. The compilation of the AI rates from
\cite{Boulmer1983,Weiner1986} is shown in figure~1. It exhibits a maximum in the AI rate constant for
Na(\textit{n}P) states near \textit{n}=12. Similar dependences on the principal quantum number were observed in
experiments with other alkali Rydberg atoms \cite{Klucharev}. Comparison of these data with the theory of
\cite{Janev1980,Mihajlov1981,DumanShmatov} has shown a remarkable disagreement of the absolute values and shapes
of experimental and theoretical \textit{n} dependences. Therefore new experiments and further development of the
theory are necessary.

\begin{figure}
\begin{center}
\epsfxsize=10 cm \epsfbox{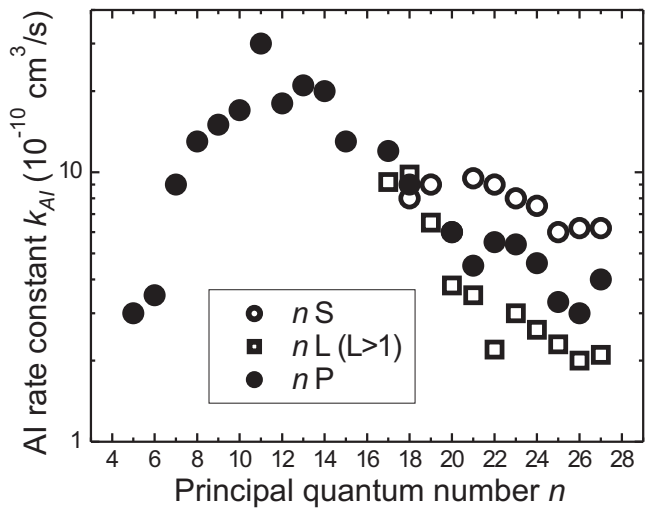} \caption{{\label{Fig1}Experimental data on the associative ionization rate
constants of $n$S, $n$P, and $nL(L>1)$ Rydberg states of sodium (from refs. \cite{Boulmer1983,Weiner1986}).}}
\end{center}
\end{figure}

One can see from figure~1 that no experimental data for sodium \textit{n}S and \textit{n}D states with $n<17$ are
available. Meanwhile, the range of $n=(8-20)$ is of particular interest, since this is where the maximum in the
ionization rate is expected. In addition, S and D states have different quantum defects and orbits of electron
motion from those of P states, and the \textit{n} dependences for S and D states may be considerably different.

One of the main difficulties in measurements of AI rates for \textit{n}S and \textit{n}D states at low \textit{n}
is related with their short lifetimes. According to the calculations of Theodosiou \cite{Theodosiou}, radiative
lifetimes of the 10S and 10D states are about 0.9 $\mu $s, while the lifetime of the 10P state is about 8 $\mu $s,
which is by an order of magnitude larger. Since one must usually provide a sufficiently long time (1-2 $\mu $s)
for the collisional interaction of Rydberg and ground-state atoms in experiments, care must be taken to
renormalize the ionization signals and thus to account for the short lifetimes.

Experimental data on effective lifetimes of most of the Rydberg states are lacking, therefore they must be
accurately calculated. Such calculations should take into account the fact that lifetimes are affected by
blackbody radiation (BBR), which is always present in the experiments. It induces transitions between neighboring
Rydberg states and shortens the effective lifetimes \cite{Gallagher}. Moreover, BBR may photoionize Rydberg atoms
and contribute Na$^{+}$ ions to the measured collisional signals. Our numerical calculations have shown (see
summary Table 1 in Section 4 below) that the rate of direct photoionization by BBR is as high as 1147 s$^{-1}$ for
the 17D state at the ambient temperature \textit{T}=300 K, so that the measured Na$^{+}$ signals in
room-temperature experiments are rather due to the BBR photoionization than due to the PI. This effect was studied
experimentally earlier in \cite{Burkhardt1984,Burkhardt1986,Spencer}, while numerical calculations of direct BBR
photoionization rates were performed in \cite{Spencer,Lehman}. Experimental measurements of BBR photoionization
rates for various \textit{n} were reported only for barium Rydberg atoms \cite{Allegrini1988}.

In order to decrease the BBR in the experiments with Rydberg atoms, special thermal shields cooled by vacuum
cryostats are often used. However, this method does not allow a precise determination of the effective temperature
and density of blackbody photons. Even with liquid nitrogen or helium cooled shielding, the blackbody photons can
penetrate into the reaction zone from the hot atomic beam source (oven) or through openings in the detection
system. Although our setup was equipped with a nitrogen cryostat, we decided to work without cooling for the sake
of exact knowledge of temperature and frequency distribution of BBR photons. Under such conditions, reliable
measurements of PI rates were impossible because the expected Na$^{+}$ signals due to PI were much smaller than
those from BBR photoionization.

At the same time, the BBR photoionization signals may be calculated or measured and serve as a reference for
measurements of the AI rates. In this case, however, the mixing of neighboring Rydberg states by BBR must also be
taken into account, because it may significantly redistribute the populations of excited states during the
interaction time \cite{Cooke,Galvez}. Note that the AI rates can still be extracted directly from the measured
Na$_{2}^{+}$ signals, since the BBR photoionization leads only to the formation of Na$^{+}$ ions and does not
affect the Na$_{2}^{+}$ signals.

\section{Experiment}

Experimental setup was similar to that used for microwave spectroscopy of Na Rydberg atoms \cite{Ryabtsev}.
Experiments were performed with a single effusive Na atomic beam in a vacuum chamber at background pressure of 5$
\cdot $10$^{-7 }$ Torr (figure~2). The temperature of the Na oven was stabilized at 635 K. The atomic beam was
formed by expansion of sodium vapor through a 2 mm dia opening in the oven at a distance of 9 cm from the reaction
zone. Collimation of the beam was achieved by a metallic shield with 1.5 mm dia aperture located 4 cm upstream
from the excitation volume. The effective diameter of the atomic beam in this volume was about 4 mm.

The metallic shield screened the most of thermal radiation of the oven. This shield and other parts of the vacuum
chamber surrounding the reaction zone were kept at room temperature, ensuring an isotropic 300 K environment. The
estimated flux of "hot" BBR photons originating from the beam source and penetrating through the collimating
aperture contributed to less than 1\% of the total BBR ionization rate in the reaction zone, and it was therefore
disregarded in the data analyses.

\begin{figure}
\begin{center}
\epsfxsize=11 cm \epsfbox{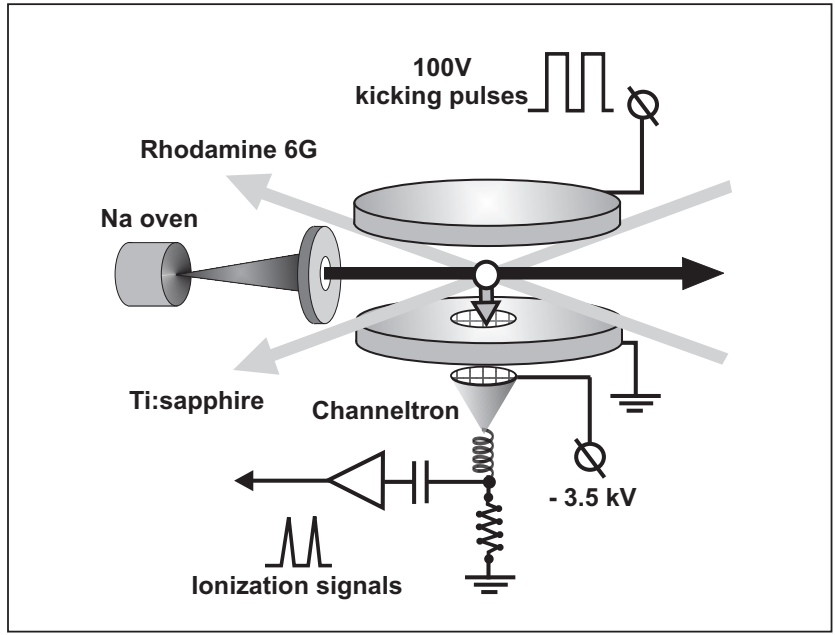} \caption{{\label{Fig2}
Experimental arrangement of the atomic and laser beams,
and the ion detection system.}}
\end{center}
\end{figure}

Sodium \textit{n}S and \textit{n}D Rydberg states were excited using the two-step scheme
3S$_{1/2}\to$3P$_{3/2}\to$\textit{n}S, \textit{n}D by radiation of two tunable lasers pulsed at 5 kHz repetition
rate. In the first step, 50 ns pulses from a Rhodamine 6G dye-laser with linewidth of 50 GHz were used. They
saturated the 3S$_{1/2}\to$3P$_{3/2}$ transition at 589 nm. The resonance fluorescence on this transition was
detected by a photomultiplier to monitor the relative changes in density of the atomic beam. In the second step,
the second harmonic of a Ti-sapphire laser was used. It yielded 50 ns pulses with 10 GHz linewidth, tunable in the
400-455 nm range. When not focused, this radiation did not saturate the 3P$_{3/2}\to$\textit{n}S,\textit{n}D
transitions. The two laser beams were crossed at a right angle in the reaction zone, both of them crossing the
atomic beam at an angle of 45$^\circ$. Laser beam profiles were spatially limited by 2 mm dia apertures at the
entrance windows of the vacuum chamber. Such configuration ensured a sufficiently small excitation volume of 2 mm
size in the central part of the atomic beam, where the spatial variation of atom density was insignificant
($<$20\%).

The ion detection system shown in figure~2 used a channeltron multiplier $\mathrm{VEU-6}$. The atomic beam passed
between two stainless-steel plates with diameter of 70 mm each, spaced by 10 mm. The plates formed a homogeneous
electric field to guide the ions from the reaction zone through the entrance window of the channeltron. The
extraction electric field pulses of 100 V/cm amplitude and 250 ns duration were applied to the upper plate. The
lower plate was grounded and had a 6 mm dia opening covered by a mesh with the transmittance of 70\%. The ions
that passed through the mesh were accelerated by the electric field of the channeltron to energies of about 3.5
keV. This energy was sufficient to detect the ions with probability close to unity.

A special microwave-spectroscopy experiment, similar to that described in \cite{Ryabtsev}, has proven that the
electric field from the channeltron very weakly penetrated into the interaction region through the fine mesh of
the lower ionization plate, as well as other stray electric fields. The measured Stark shift of the microwave
transition 37S$\to$37P was about 4 MHz, corresponding to the $<$0.3 V/cm stray electric field. This field had no
effect on our measurements or on Rydberg states under study.

Single ion output pulses of the channeltron were amplified, discriminated, and registered by two gated counters.
The measurements were performed in the pulse counting regime, keeping the frequencies of the detected ion signals
much lower (0.2-1~kHz) than the 5 kHz repetition rate of laser pulses. On average, less than one ion per laser
shot was detected. The measured frequencies of ion signals were determined as the total number of ions detected
during the measurement time of 10 s, i.e., signals were counted for 50000 laser pulses. In order to ensure the
single-ion counting regime, the intensity of the laser driving the second excitation step was attenuated with
calibrated neutral density filters by a factor of 10 to 100.

The present study concentrated on measurements of relative dependence of ionization rate constants on the
principal quantum number \textit{n} of Rydberg states, and it did not require precise knowledge of absolute values
of the number density \textit{n}$_{3S}$ of the ground-state Na(3S) atoms in the beam. It was therefore calculated
from the geometry of the effusive beam and the Nesmeyanov's formula \cite{Nesmeyanov} relating temperature and
pressure of saturated Na vapor in the beam source. This formula coincides to within better than 50\% with other
available data, for example, \cite{PhysicalValues}. Monitoring of the fluorescence on the saturated resonance
transition showed that the atomic number density was almost constant during the experiments. The temperature of
the oven was measured by a calibrated thermocouple with an accuracy of $ \pm $2 K. Taking into account all the
uncertainties, we estimate the absolute number density of atoms in the reaction zone as \textit{n}$_{3S}$=(2$ \pm
$0.7)$ \times $10$^{10}$ cm$^{-3}$ at the oven temperature of \textit{T}=(635$ \pm $2) K.

The time sequence of excitation and detection pulses is illustrated in figure~3. Two identical electric field
pulses with the 100 V/cm amplitude and 250 ns duration (figure~3(b)) were applied to the repelling plate after
each laser excitation pulse (figure~3(a)).

The first electric pulse was applied immediately after the laser pulse to remove the atomic and molecular ions
produced during the laser pulse. These ions were always observed in our experiments, even at non-resonant
excitation by the first-step (\textit{yellow}) laser only. Their origin was investigated in many earlier studies,
which identified them to be due to mutiphoton ionization of Na$_{2}$ dimers by yellow laser \cite{BoulmerDimer,
Burkhardt1985} (dimers are always present in an effusive beam at an amount of $\sim $(0.1-1)\%), due to
photoassociative ionization of Na(3P) atoms by yellow laser \cite{Meijer,Dulieu,Blange}, and due to
photoionization of Rydberg Na(\textit{nL}) atoms by the resonant radiation of yellow and blue lasers. These ions
represent an undesirable background to the collisional-ionization signals from Rydberg atoms, and therefore the
first electric pulse was required to clean up the reaction zone from any ions created during the laser excitation.

The second electric pulse extracted to the channeltron those ions, which were produced in the time interval
between \textit{t}$_{1}$=0.3 $\mu $s and \textit{t}$_{2}$=2.1 $\mu $s after the laser excitation pulse. These ions
corresponded to the collisional and BBR photoionization of Rydberg atoms. In the mass spectrum detected by the
channeltron, the signals of the atomic Na$^{+}$ and the molecular Na$_{2}^{+}$ ions were separated by 0.6 $\mu s$
and thus well resolved (figure~3(c)). Gated pulse counters registered the signals of the atomic and molecular ions
independently (figure~3(d)).

\begin{figure}
\begin{center}
\epsfxsize=10 cm
\epsfbox{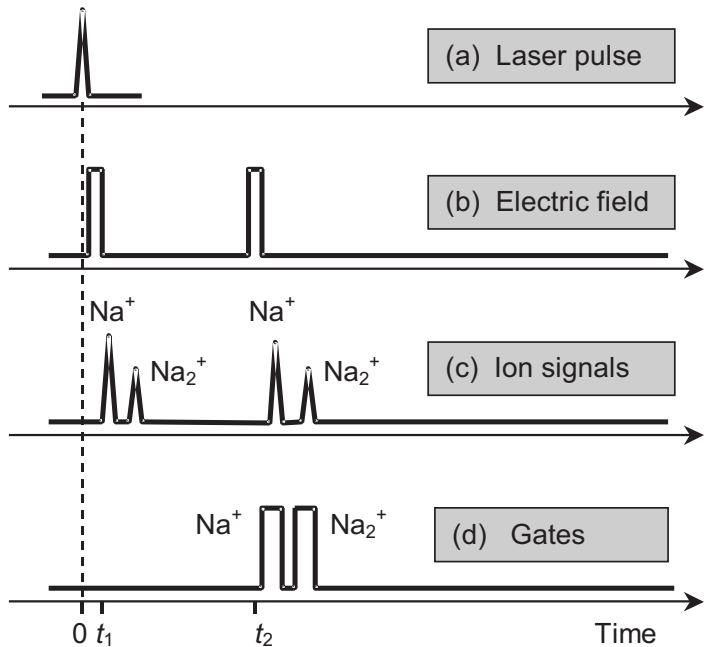} \caption{{\label{Fig3}Timing diagram of signals: (a)
laser excitation pulse; (b) 100 V/cm electric field pulses for
ion extraction; (c) ion signals; (d) counter gates for measurement
of $\mbox{Na}^+$ and $\mbox{Na}_2^+$ signals.}}
\end{center}
\end{figure}

\section{Determination of collisional rate constants}

\subsection{Rate equations}

The registered Na$^{+}$ and Na$_{2}^{+}$ ion signals
result from ionization occurring in the reaction zone during the time
interval (\textit{t}$_{2}$-\textit{t}$_{1}$)=1.8 $\mu $s between the two
extraction electric field pulses. This time is comparable with the lifetimes
of Rydberg states; therefore, time evolution of ionization processes must be
analyzed. The main processes leading to the formation of Na$^{+}$
ions are Penning ionization and photoionization by BBR. The
Na$_{2}^{+}$ ions can be created only in the associative
ionization. Contribution of the collisions with background gases can be
safely disregarded. We have verified experimentally that the variation of
background pressure within the range of 5$\cdot $10$^{-7}  \le
$ \textit{P} $ \le $ 1$ \cdot $10$^{-6}$ Torr did not affect the measured
Na$^{+}$ and Na$_{2}^{+}$ signals by more than 5\%.
Under such conditions, the rate equations describing the evolution of the
numbers of Na$^{+}$ and Na$_{2}^{+}$ ions following the
laser excitation at time \textit{t}=0 are

\begin{equation}
\label{eq1} \left\{{{\begin{array}{*{20}l} {\frac{\mathstrut {\mbox{\textit{d}Na}^{ +} \left(
{t} \right)}}{{\mbox{\textit{dt}}}} = k_{PI} N_{nL} \left( {t} \right)\,n_{3S} +
W_{BBR} N_{nL} \left( {t} \right);}\\ \\

{\frac{{\mbox{\textit{d}Na}_{{2}}^{ +} \left(
{t} \right)}}{{\mbox{\textit{dt}}}} = k_{AI} N_{nL} \left( {t} \right)\,\,n_{3S}.}
\\
\end{array}} } \right.
\end{equation}

\noindent Here, \textit{N}$_{nL}$(\textit{t}) is the time-dependent number of Na(\textit{nL}) Rydberg atoms in the
reaction zone, \textit{n}$_{3S}\approx$2$ \cdot $10$^{10}$ cm$^{-3}$ is the number density of ground state atoms,
\textit{k}$_{AI}$ and \textit{k}$_{PI}$ are the rate constants of associative and Penning ionization in
Na(\textit{nL})+Na(3S) collisions respectively.

The function \textit{N}$_{nL}$(\textit{t}) evolves in time as

\begin{equation}
\label{eq2}
N_{nL} \left( {t} \right) \approx N_{nL} \left( {0} \right)\,\,\rm exp\mit \left[ { -
t/\tau _{eff}}  \right].
\end{equation}

The initial number of Rydberg atoms, \textit{N}$_{nL}$(0), created during
the laser excitation, can be written as

\begin{equation}
\label{eq3}
N_{nL} \left( {0} \right) = N_{3P} W\left( {3P_{3/2} \to nL} \right),
\end{equation}

\noindent where \textit{N}$_{3P}$ is the average number of atoms in the 3P$_{3/2}$
state during the yellow-laser
pulse, and $W\left( {3P_{3/2} \to nL} \right)$ is the probability of excitation of the Na(3P$_{3/2}$)
atoms to the \textit{nL} state by a single blue-laser shot.

The effective lifetime $\tau _{eff}$ describing the decay of Rydberg states in equation~(\ref{eq2}) is determined
by the spontaneous lifetime and the rate of other processes depleting the laser excited Rydberg state. These
include BBR induced transitions between Rydberg states, BBR induced photoionization, and collisional quenching.
The results of our numeric calculations of the rates of these processes will be given in Section 4 and summarized
in Table 1.

The depletion of Rydberg states with \textit{n}=8-20 by collisional ionization is negligible at atom densities
used in the present experiment. We find from figure~1 that the rate of associative ionization,
\textit{k}$_{AI}$\textit{n}$_{3S}$, does not exceed 50 s$^{-1}$ and is therefore much smaller than the spontaneous
decay rates, which range from 10$^{5}$ to 10$^{6}$ s$^{-1}$ for the studied Rydberg states. The rate of PI,
\textit{k}$_{PI}$\textit{n}$_{3S}$, is expected to be below 10 s$^{-1}$ for \textit{n}$\sim $20, and close to zero
for lower \textit{n}. Comparing the PI rate with the direct BBR photoionization rate \textit{W}$_{BBR}$ (Table 1),
one can see that Na$^{+}$ ions are produced mainly in the BBR photoionization. As will be shown below, this
seemingly undesirable background ionization process can be favorably exploited for the determination of absolute
AI rate constants.

With the above considerations in mind, the solution of equations~(\ref{eq1}) can be written as

\begin{equation}
\label{eq4}
\left\{ {{\begin{array}{*{20}c}
 {\mbox{Na}^{ +}  = N_{nL} \left( {0} \right)\;W_{BBR} \;t_{eff}}  \\
 {\mbox{Na}_{2}^{ +}  = N_{nL} \left( {0} \right)k_{AI} n_{3S} \;t_{eff}}  \\
\end{array}} } \right.
\end{equation}

\noindent
where \textit{t}$_{eff}$ is the effective time of interaction that takes
into account the short radiative lifetimes of Rydberg states:

\begin{equation}
\label{eq5}
t_{eff} = \tau _{eff} \left[ {\rm exp \mit \left( { - t_{1} /\tau _{eff}}  \right) -
\rm exp \mit\left( { - t_{2} /\tau _{eff}}  \right)} \right].
\end{equation}
 \noindent
Here, \textit{t}$_{1}$ and \textit{t}$_{2}$ are the time moments marking the end of the first and the beginning of
the second extraction electric field pulses, respectively (see figure~3(b)).

\subsection{Method of measurements}

Equations(\ref{eq4}) can be used for the direct measurement of \textit{k}$_{AI}$ and \textit{W}$_{BBR}$ values,
provided that \textit{N}$_{nL}$(0) is known. The only reliable method to measure \textit{N}$_{nL}$(0) is the
Selective Field Ionization (SFI) technique \cite{SFI}.  Rydberg atoms ionize in an electric field with probability
close to unity if the field strength has reached the critical value \textit{E}$_{c}$. The latter depends strongly
on the effective quantum number \textit{n}$_{eff}$=(\textit{n}-$\delta _{L}$), where $\delta _{L}$ is the quantum
defect (1.348 for \textit{n}S states and 0.015 for \textit{n}D states):

\begin{equation}
\label{eq6}
E_{c} \approx 3.2 \cdot 10^{8}n_{eff}^{ - 4} \quad \left( {V/cm} \right)
\end{equation}

\noindent Unfortunately, it is difficult to apply the SFI method for Rydberg states with low \textit{n}, since
they require too strong electric field ($\sim $30 kV/cm for \textit{n}$\sim $10). In the present study we were
interested mainly in relative measurements of \textit{k}$_{AI}$ for various \textit{n}. Therefore we could use a
normalization procedure for \textit{N}$_{nL}$(0) based on numerically calculated excitation probabilities $W\left(
{3P_{3/2} \to nL} \right)$. Since the 3S$_{1/2} \to $3P$_{3/2}$ transition was saturated, \textit{N}$_{nL}$(0)
was dependent only on the respective transition moments and power of the blue laser. In the absence of saturation at
the second excitation step (this was the case in our experiments), the probability of excitation of Rydberg
states from the 3P$_{3/2}$ state can be written as

\begin{equation}
\label{eq7}
W\left( {3P_{3/2} \to nL} \right) = C_{L} \cdot P_{b} \cdot R^{2}\left(
{3P_{3/2} \to nL} \right),
\end{equation}

\noindent where \textit{P}$_{b}$ is the power of the blue laser, $R\left( {3P_{3/2} \to nL} \right)$ is the radial
part of the transition dipole moment, and \textit{C}$_{L}$ is a normalization constant, which depends on
\textit{L} and is proportional to the square of angular part of the matrix element. The value of $W\left(
{3P_{3/2} \to nL} \right)$ falls as $n_{eff}^{ - 3} $ for high Rydberg states, but for the states with
\textit{n}$\sim $10 this scaling law does not work quite well. This was established in our numeric calculations of
$R\left( {3P_{3/2} \to nL} \right)$ for the 3P$_{3/2} \to $\textit{n}S, \textit{n}D transitions, therefore we used
the calculated values instead of the above scaling law for the evaluation of the experimental data.

In order to compare the absolute signals due to BBR and collisional
ionization of S states with those of D states, it is necessary to know also the ratio
\textit{C}$_{D}$/\textit{C}$_{S}$. The analysis of angular parts of the
transition matrix elements, taking into account the hyperfine structure,
showed that for excitation with linearly polarized light in the first and
the second excitation steps, the ratio \textit{C}$_{D}$/\textit{C}$_{S}$ may
vary from approximately 1.6 (if there is no collisional, radiative, or
magnetic field mixing of the magnetic sublevels) to 2 (if the level mixing
is complete). For excitation by non-polarized light, the ratio always equals
to 2 regardless the degree of level mixing. Finally, we find that the ratio
$W\left( {3P_{3/2} \to nD} \right)/W\left( {3P_{3/2} \to nS} \right)$ may
vary between the values of 3.5 and 5.

In principle, one could normalize the ion signals measured for different \textit{nL} states using the calculated
probabilities $W\left( {3P_{3/2} \to nL} \right)$ and measuring only the power \textit{P}$_{b}$ of the blue laser
radiation in equation~(\ref{eq7}). However, the applicability of such normalization may be complicated by
technical imperfections of the blue laser. Since the linewidth of this laser (10 GHz) was much larger than the
widths of the absorption profiles at the second excitation step ($\sim $500 MHz Doppler broadening), variations of
the spectral density of laser radiation could affect the probability of excitation even if \textit{P}$_{b}$ would
be kept constant. Therefore we had to verify experimentally the applicability of normalization by
equation~(\ref{eq7}). As discussed above, the only reliable way to measure the number of Rydberg atoms was to
apply the SFI technique. For this purpose, we built a high-voltage generator yielding pulses with rise time of 1
$\mu s$ and amplitude of up to 8 kV. This allowed us to field-ionize Rydberg states with \textit{n$ \ge $}17. The
SFI signals were detected at a 1 $\mu s$ delay with respect to the laser pulse, i.e., the measured SFI signal was:

\begin{equation}
\label{eq8}
S_{SFI} \sim N_{nL}({0})\exp( {- 1\;\mu s/\tau _{eff}}).
\end{equation}

Equation(\ref{eq8}) was used to derive \textit{N}$_{nL}$(0) from the measured SFI signals and the calculated
values $\tau _{eff}$ of Table 1. Figure 4 shows the measured \textit{N}$_{nL}$(0) dependences on the principal
quantum number \textit{n} for \textit{n}S and \textit{n}D states. These data are normalized on \textit{P}$_{b}$,
because it varied as blue laser frequency was tuned to resonance with different \textit{nL} states. The solid
curves are approximations made using equation~(\ref{eq7}). It is seen that experimental points have noticeable
deviations from theory although the general trend is correct. These deviations may be explained by the variations
of spectral density of blue laser light. We concluded that equation~(\ref{eq7}) can be used for the normalization
of \textit{N}$_{nL}$(0), but at price of limited accuracy. We also find from figure~4 that on average the ratio
$W$(3P$_{3/2} \to $\textit{n}D)/$W$(3P$_{3/2} \to $\textit{n}S) was close to 3.5. Hence, no considerable mixing of
the magnetic sublevels took place during the laser excitation, and the ratio \textit{C}$_{D}$/\textit{C}$_{S}$ was
close to 1.6.

\begin{figure}
\begin{center}
\epsfxsize=15 cm
\epsfbox{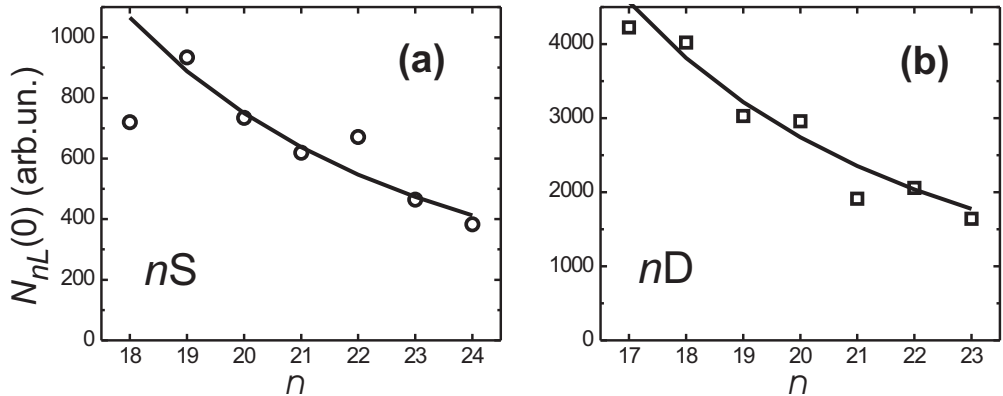} \caption{{\label{Fig4}Relative probabilities of laser excitation of sodium
Rydberg states: (a) $n$S states; (b) $n$D states. Open circles and squares - experiment,
solid curves - theory.}}
\end{center}
\end{figure}
Since SFI technique was not applicable for the direct determination of \textit{N}$_{nL}$(0) for \textit{n}=8-17,
the use of equation~(\ref{eq7}) was  somewhat inadequate in our experiment. Therefore ways had to be found to
eliminate this value from the determination of rate constants. This can be done by measuring the ratio \textit{R}
of atomic and molecular signals derived from equations~(\ref{eq4}):

\begin{equation}
\label{eq9}
R = \frac{{\mbox{Na}_{2}^{ +} } }{{\mbox{Na}^{ +} }} = \frac{{k_{AI} n_{3S}} }{{W_{BBR}
}}.
\end{equation}

This ratio does not depend on the values of \textit{N}$_{nL}$(0), $\tau
_{eff}$ and \textit{t}$_{eff}$ . Thus, the rate constant of the AI process
can be directly obtained from the measured ratio of the
Na$_{2}^{+}$ and Na$^{+}$ signals:

\begin{equation}
\label{eq10}
k_{AI} = \frac{{\mbox{Na}_{2}^{ +} } }{{\mbox{Na}^{ +} }} \cdot \frac{{W_{BBR}} }{{n_{3S}
}}.
\end{equation}

The BBR ionization rates \textit{W}$_{BBR}$ have thus become one of the key values necessary for the determination
of the AI rate constants. The accuracy with which the \textit{W}$_{BBR}$ values are known directly affect the
accuracy of the experimental \textit{k}$_{AI}$ values obtained using equation~(\ref{eq10}). Therefore, the process
of BBR ionization must be examined in all details.

\section{BBR ionization rates}

\subsection{Direct BBR photoionization and effective lifetimes}

First, we have numerically calculated the effective lifetimes $\tau _{eff}$ and the rates
\textit{W}$_{BBR}$\textit{} of direct BBR photoionization of \textit{n}S and \textit{n}D Rydberg states of Na with
\textit{n}=8-20 at the room temperature of 300 K. The BBR photoionization rates were calculated using the formula
(here and below the atomic units are used) \cite{Spencer}:
\begin{equation}
\label{eq11}
W_{BBR} = c\int\limits_{\omega _{nL}} ^{\infty}  {\sigma _{\omega}  \;\rho
_{\omega}  d\omega}  ,
\end{equation}

\noindent where \textit{c} is the speed of light, $\omega _{nL} = 1/\left( {2n_{eff}^{2}}  \right)$ is the
threshold frequency of the \textit{nL} Rydberg state, and $\sigma _{\omega}  $ is the photoionization
cross-section at frequency $\omega $. The volume density $\rho _{\omega}  $ of BBR photons at temperature
\textit{T} is given by the Planck distribution:
\begin{equation}
\label{eq12}
\rho _{\omega}  = \frac{{\omega ^{2}}}{{\pi ^{2}c^{3}\left[ {e^{\omega
/\left( {kT} \right)} - 1} \right]}},
\end{equation}
\noindent where \textit{k} is the Boltzmann constant. For isotropic and non-polarized thermal radiation field,
$\sigma _{\omega}  $ is determined by the radial matrix elements $R\left( {nL \to E,L \pm 1} \right)$ of dipole
transitions from discrete \textit{nL} Rydberg states to continuum states with $L'=(L \pm 1)$ and photoelectron
energy \textit{E}:
\begin{equation}
\label{eq13}
\sigma _{\omega}  = \frac{{4\pi ^{2}\omega} }{{3c\left( {2L + 1}
\right)}}\sum\limits_{L' = L \pm 1} {L_{max} R^{2}\left( {nL \to EL \pm 1}
\right)} ,
\end{equation}

\noindent where \textit{L}$_{max}$ is the largest of $L$ and $L'$. The values of $R\left( {nL \to E,L \pm 1}
\right)$ were calculated numerically using the semi-classical formulae of Dyachkov and Pankratov
\cite{DyachkovPankratov}. Their method gives orthogonal and normalized continuum wavefunctions, and the calculated
photoionization cross-sections were found to be in good agreement with the more sophisticated quantum-mechanical
calculations for \textit{n}S states by Aymar \cite{Aymar}. Other theoretical methods developed in
\cite{Zimmerman,BezuglovBorodin} give close results. High-precision values of quantum defects of Dyubko et al
\cite{Dyubko} were used in our calculations.

In order to determine the effective lifetimes of $nL$ states, the radial matrix elements $R\left( {nL \to n'L'}
\right)$ of all dipole transitions to other $n'L'$ states with $L'=L \pm 1$) were also calculated using the
quasi-classical formulae of \cite{DyachkovPankratov}. The rate of a specific BBR-induced transition between the
states $nL$ and $n'L'$ are given by:

\begin{equation}
\label{eq14}
W\left( {nL \to n'L'} \right) = \frac{{4}}{{3c^{3}}}\frac{{L_{max}} }{{2L +
1}}R^{2}\left( {nL \to n'L'} \right)\frac{{\omega _{nn'}^{3}} }{{e^{\omega
_{nn'} /\left( {kT} \right)} - 1}}
\end{equation}

\noindent
where $\omega _{nn'} =1/(2n^{2})-1/(2n'^{2})$ is the
frequency of the transition. The total decay rates were obtained by summing the
probabilities of all spontaneous and  BBR-induced transitions. The results
of numerical calculations of \textit{W}$_{BBR}$ and effective lifetimes at
\textit{T}=300K are presented in Table 1.

\begin{table}
\centering \caption{ Effective lifetimes $\tau_{eff} (\mu s)$ and BBR induced transition rates of \textit{n}S and
\textit{n}D states of sodium for the surrounding environment at \textit{T}=300 K. $W_{BBR}$ - direct
photoionization rates of laser excited Rydberg states; $W_{R}$ - ionization rates by the second extraction field
pulse of states with $n'>40$, which are populated in BBR induced transitions from the initial laser excited
levels; $W_{mix}$ - rate of BBR induced ionization of $n'L'$ levels populated by spontaneous decay and BBR-induced
$nL$ mixing; $W^{tot}_{BBR}$ - total theoretical BBR ionization rate;  $W^{exp}_{BBR}$ - experimental BBR induced
ionization rate.}
\small
\begin{tabular*}{\textwidth}{@{\extracolsep{\fill}}|c|c|c|c|c|c|c|}  \hline
State & $\tau_{eff} (\mu s)$ & $W_{BBR} (s^{-1})$ & $W_{R} (s^{-1})$ & $W_{mix} (s^{-1})$&$W^{tot}_{BBR}
(s^{-1})$& $W^{exp}_{BBR} (s^{-1})$\\ \hline 8S&  0.395&   0,059&    0.022&0.404&0.485 & \\ \hline 9S&  0.581&
0.823&    0.316&3.62&4.76&6$\pm$2\\ \hline 10S& 0.812&   4.555&   1.78&13.7&20.0&25$\pm$2\\ \hline 11S& 1.095&
14.48&   5.76&30.9&51.2&47$\pm$3\\ \hline 12S& 1.431&   32.36&   13.2&52.5&98.0 &93$\pm$10\\ \hline 13S& 1.824&
57.42&   23.9&73.9&155 &141$\pm$16 \\ \hline 14S& 2.278&   87.07&   37.4&91.8&216& 204$\pm$23\\ \hline 15S& 2.795&
118.3&   52.6&104& 275& 268$\pm$24\\ \hline 16S& 3.376&   148.8&   68.6&111& 329& 367$\pm$33 \\ \hline 17S& 4.024&
176.3&   85.0&114& 375& 499$\pm$55\\ \hline 18S& 4.731&   200.9&   101& 113& 415& 614$\pm$59\\ \hline 19S& 5.518&
221.5&   117& 109& 448& 741$\pm$66\\ \hline 20S& 6.378&   238.1&   133& 104& 475& 1007$\pm$110\\ \hline
\multicolumn{7}{|c|}{}\\ \hline 8D&  0.474&  12.70&   4.19&  5.7&     22.6&    16$\pm$ 7\\ \hline 9D&  0.665&
52.76&   17.4&  15.1&    85.3&    73$\pm$14\\ \hline 10D& 0.909&   138.0&   45.6&  28.4&    212.0&   188$\pm$25\\
\hline 11D& 1.189&   268.8&   89.4&  43.8&    401.9&   369$\pm$38\\ \hline 12D& 1.524&   424.3&   145&   58.0&
627.2&   633$\pm$67\\ \hline 13D& 1.905&   603.4&   207&   69.8 &   879.9&   830$\pm$69\\ \hline 14D& 2.341&
770.4&   270&   78.3 &   1119 &   1125$\pm$128\\ \hline 15D& 2.835&   920.0&   331&   83.4 &   1335 &
1309$\pm$70\\ \hline 16D& 3.387&   1046 &   389&   85.4 &   1521 &   1420$\pm$82\\ \hline 17D& 4.001&   1147 &
443&   85.1 &   1675 &   1630$\pm$67\\ \hline 18D& 4.680&   1225 &   492&   82.9 &   1800 &   1804$\pm$88\\ \hline
19D& 5.423&   1280 &   538&   79.3 &   1897 &   1978$\pm$96\\ \hline 20D& 6.235&   1317 &   580&   74.9 &   1972 &
2036$\pm$60\\ \hline
\end{tabular*}
\end{table}
\normalsize

\subsection{BBR-induced mixing of Rydberg states}

Earlier studies of BBR-induced processes have revealed that BBR causes not only direct photoionization to the
continuum \cite{Spencer}, but also induces transitions in the discrete spectrum \cite{Cooke,Galvez}. In our case,
several discrete \textit{n}P and \textit{n}F levels are populated from the initially excited \textit{n}S and
\textit{n}D states during the interaction time. For example, after laser excitation of the 16S state, the
BBR-induced transitions populate the neighboring $n'$P states (figure~5(a)). These states have significantly
larger \textit{W}$_{BBR}$ values than that of the 16S state, so that population transfer to the $n'$P states may
affect the resultant BBR photoionization signal. As an example, the calculated rates of spontaneous and
BBR-induced transitions from the 16S and the 16D states to the $n'$P states are shown in figures~5(b) and 5(c).

\begin{figure}
\begin{center}
\epsfxsize=15 cm \epsfbox{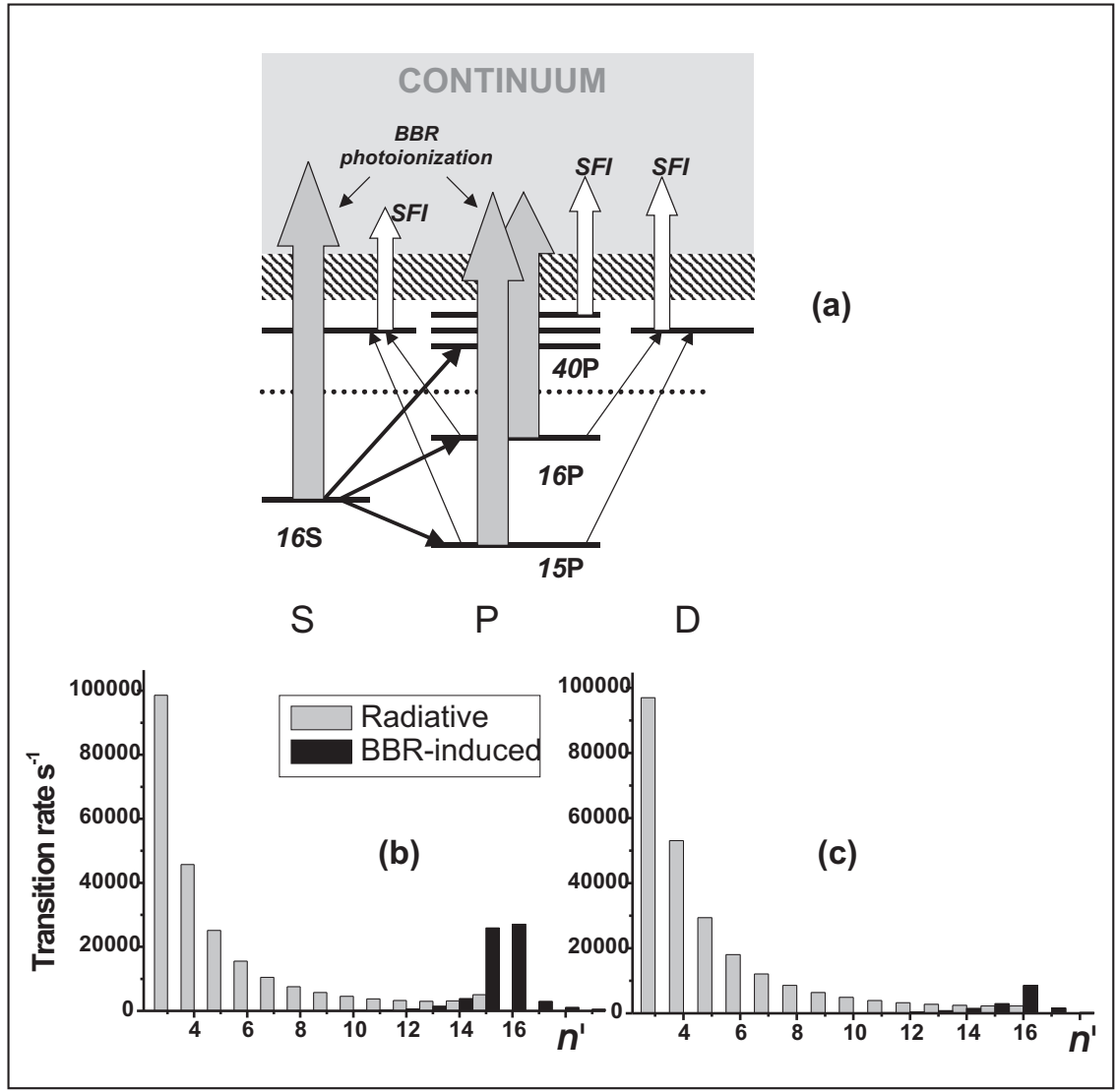} \caption{{\label{Fig5}(a) Schematic illustration of BBR induced and field
ionization processes upon excitation of the 16S state. They include redistribution of population over other $n'L'$
Rydberg states due to spontaneous and BBR induced transitions from the initial state. (b) Calculated spontaneous
and BBR induced transition rates from the initial 16S state to other $n'$P states. (c) Calculated spontaneous and
BBR induced transition rates from the initial 16D state to other $n'$P states.}}
\end{center}
\end{figure}

Another feature is that BBR also induces transitions to high Rydberg states (figure~5(a)), which are field ionized
by the second electric field pulse extracting the ions to channeltron for detection. This 100 V/cm pulse ionizes
Rydberg states with $n \ge $ 40, according to equation~(\ref{eq6}). Numerical calculations showed that high lying
Rydberg states are populated at rates comparable with the direct photoionization rate ${W}_{BBR}$ given in Table
1. Since all the states above $n_{c}$ = 40 are field ionized, all BBR populated states with $n' \ge $40 will
contribute to the detected Na$^{+}$ ion signal. The total rate $W_{R}$ of BBR transitions to all Rydberg states
with $n' \ge 40$ was calculated by summing the individual $nL \to n'L'$ contributions given by
equation~(\ref{eq14}):

\begin{equation}
\label{eq15}
W_{R} = \,\sum\limits_{n' \ge 40} {\,\sum\limits_{L' = L \pm 1} {W\left( {nL
\to n'L'} \right)}}  .
\end{equation}

The calculated $W_{R}$ values are given in Table 1. Note, that
these ionization rates amount to between 30 to 50 \% of the direct
photoionization rate values, $W_{BBR}$, and they obviously cannot
be ignored.

Another kind of processes that may also lead to the redistribution of populations between Rydberg states is $nL$
mixing in collisions with the abundant ground state atoms. However, the estimated rates of this process for a beam
of density  $n_{3S} \approx 2 \times 10^{10}$ cm$^{-3}$ produced by a source at 635 K were found to be much lower
than the rates of spontaneous and BBR-induced transitions. Therefore the contribution of collisional $nL$ mixing
can be safely disregarded.

\subsection{Total BBR ionization rates}

When the rates of redistribution of population over $n'L'$ Rydberg states following laser excitation of a single
\textit{nL} Rydberg level are known, the total ionization rate leading to the production of Na$^{+}$ ions
can be calculated. Assume that some \textit{n}S state is initially laser excited. If the rate of Penning
ionization is negligible, the production of Na$^{+}$ ions is described by the rate equation

\begin{equation}
\label{eq16}
\frac{{d\,\mbox{Na}^{ +} \left( {t} \right)}}{{d\,t}} = W_{BBR}^{nS} N_{nS} \left(
{t} \right) + \sum\limits_{n'} {W_{BBR}^{n'P} N_{n'P} \left( {t} \right)} .
\end{equation}

\noindent Here, $W_{BBR}^{nS} = W_{BBR} + W_{R} $ is the total BBR ionization rate of the laser excited
\textit{n}S state, and it consists of two contributions: the direct BBR photoionization rate $W_{BBR}$ of the
\textit{n}S state given by equation~(\ref{eq11}), and the rate $W_{R}$ of equation~(\ref{eq15}) for BBR induced
excitation of states with  $n \ge $ 40, which are subsequently field ionized by the second extraction field pulse.
The main contributions to the sum in equation~(\ref{eq16}) are those of $n'$P states with $n'=n,(n-1)$, which are
the nearest neighbors to the initial \textit{n}S state (see figures~5(b) and 5(c)). The time evolution of the
number of $n'$P atoms is described as follows:

\begin{equation}
\label{eq17}
\frac{{d\,N_{n'P} \left( {t} \right)}}{{d\,t}} = W\left( {nS \to n'P}
\right)N_{nS} \left( {t} \right).
\end{equation}

\noindent Solving equations (\ref{eq16}) and (\ref{eq17}), we obtained the total effective ionization rate of the
Na(\textit{n}S) atoms:

\begin{equation}
\label{eq18}
W_{BBR}^{tot} = W_{BBR} + W_{R} + W_{mix},
\end{equation}

\noindent
where $W_{mix}$ is a contribution from the spontaneous and
BBR-induced mixing of neighboring states:

\begin{equation}
\label{eq19}
W_{mix} = \sum\limits_{n'} {W\left( {nS \to n'P} \right)\;W_{BBR}^{n'P}}
\left( {\frac{{t_{2} - t_{1}} }{{t_{eff}} } - 1} \right)\tau _{eff}.
\end{equation}

\noindent Here $t_{eff}$ is the effective interaction time given by equation~(\ref{eq5}), and $\tau _{eff}$ is the
effective lifetime of the laser excited \textit{n}S state. Similar equations were derived for \textit{n}D states,
except that $W_{mix}$ accounted for the transitions to both $n'$P and $n'$F states. Results of numerical
calculations using the semi-classical formulae for matrix elements from \cite{DyachkovPankratov} are summarized in
Table 1.

\subsection{Comparison of theory and experiment for BBR ionization rates}

Experimental and theoretical data on the total BBR photoionization rates are compared in figure~6. The solid
curves are the theoretical values of $W_{BBR}^{tot} $ from Table 1. The squares and circles are the corresponding
experimental values $W_{BBR}^{exp} $ obtained using equations~(\ref{eq4}) and (\ref{eq7}) (in equation~(\ref{eq4})
the value of \textit{W}$_{BBR}$ was replaced with $W_{BBR}^{tot} $). Experimental data were averaged over 5
measurements. The normalization coefficient \textit{C}$_{D}$\textit{} in equation~(\ref{eq7}) was the only
parameter whose absolute value was adjusted to fit the experiment to the theory. A good agreement between the
experimental and theoretical data was found for \textit{n}D states (figure~6(b)). However, the data for
\textit{n}S states, obtained with the ratio \textit{C}$_{D}$/\textit{C}$_{S}$=1.6  measured earlier, exhibit
considerable discrepancies for states with higher \textit{n} (figure~6(a)), while a better agreement for states
with lower \textit{n} is seen. The values of $W_{BBR}^{exp} $ exceed the values of $W_{BBR}^{tot} $ by 2.1 times
for \textit{n}=20, and the shape of the experimental \textit{n} dependence is significantly different from the
theoretical one.

\begin{figure}
\begin{center}
\epsfxsize=15 cm
\epsfbox{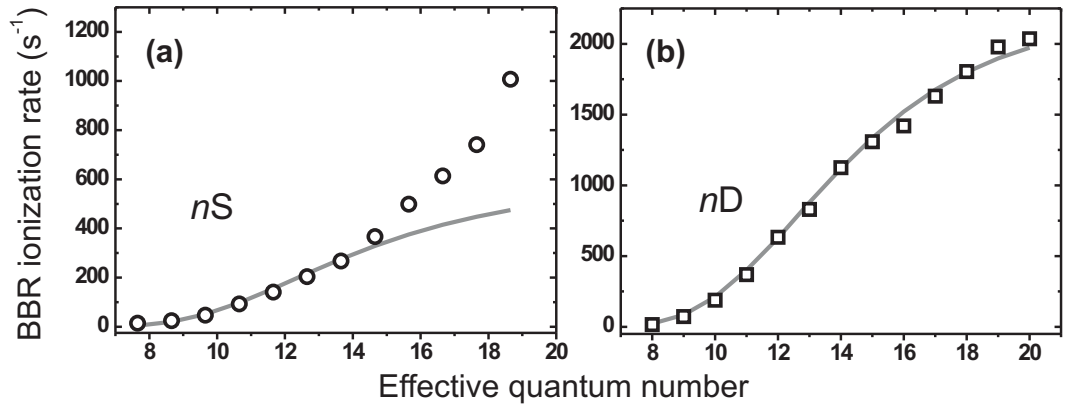} \caption{{\label{Fig6}Total BBR induced ionization rates for (a) $n$S states and
(b) $n$D states. Open circles and squares - experiment, solid lines - theory.}}
\end{center}
\end{figure}

One possible explanation of such anomaly for \textit{n}S states is related to their specific orbit that penetrates
into the atomic core. The penetration causes a strong interaction between the Rydberg electron and the core, e.g.,
due to core polarization \cite{Aymar}. This results in a large quantum defect and in a Cooper minimum in the
photoionization cross-sections. Each \textit{n}S state has its individual photoionization cross-section and
position of the Cooper minimum. The accuracy of theoretical frequency-dependent cross-section values affects the
total BBR ionization rate when integration over the Planck distribution in equation~(\ref{eq11}) is performed. It
is likely that the accuracy of theoretical models for \textit{n}S states is affected by difficulties to account
for the core penetration. This assumption is supported by good agreement of theory and experiment for the
hydrogen-like \textit{n}D states, which have a small quantum defect and an almost non-penetrating orbit.

We conclude that our theoretical calculations of $W_{BBR}^{tot}$ are correct to better than 10\% for \textit{n}D
states and lower \textit{n}S states, and probably incorrect for higher \textit{n}S states. New more accurate
calculations of S state ionization rates are required if equation~(\ref{eq10}) is to be used for the determination
of absolute AI rate constants \textit{k}$_{AI}$ for \textit{n}S states. However, such calculations require
development of new approaches that are not yet available for Rydberg states. Until new theoretical data become
available, we rely on our experimental data for \textit{n}S states instead of theoretical BBR ionization rates.
The experimental values of $W_{BBR}^{exp} $ are summarized in the last column of Table 1. These are our
recommended values of the BBR ionization rates at 300 K, which should be used in equation~(\ref{eq10}).

\section{Experimental results for associative ionization rate constants}

\subsection{Ratio of molecular and atomic ion signals}

Figure 7 shows the ratio \textit{R} of equation~(\ref{eq9}) averaged over eight independent measurements at the
density of atoms in the atomic beam \textit{n}$_{3S}$=2$ \times $10$^{10}$ cm$^{-3}$. For the lowest states, the
values of \textit{R} reach 2.5 and 1.4 for \textit{n}S and \textit{n}D states, respectively. In both cases,
\textit{R} drops off rapidly and asymptotically approaches the value of 0.06 for highest \textit{n}.

\begin{figure}
\begin{center}
\epsfxsize=15 cm \epsfbox{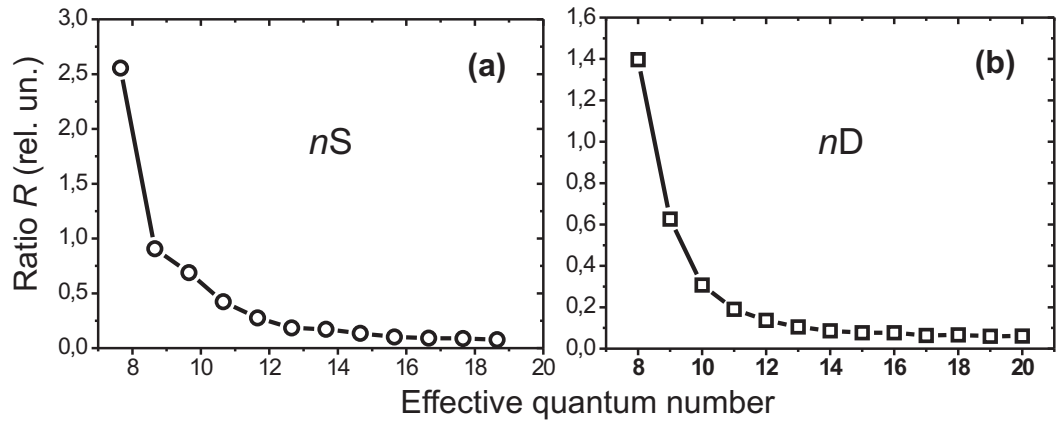} \caption{{\label{Fig7}Experimental ratio $R$ of Na$_2^+$ and Na$^+$ ion
signals as a function of the effective quantum number $n_{eff}$ for (a) $n$S states and (b) $n$D states.}}
\end{center}
\end{figure}

Such nearly constant \textit{n} dependence of \textit{R} at higher \textit{n} was unexpected. In order to verify
whether the origin of the measured signal is collisional, the dependence of \textit{R} on the atomic beam density
was measured for low (9D and 10S) and high (19S and 18D) Rydberg states (figure~8). For the 9D and 10S states,
\textit{R} has a linear dependence on \textit{n}$_{3S}$, and hence the Na$_{2}^{+}$ signal has a quadratic one.
For 18D and 19S states, \textit{R} is almost independent of \textit{n}$_{3S}$, hence the dependence of
Na$_{2}^{+}$ is linear. This means that for high \textit{n} the apparent Na$_{2}^{+}$ signal was not due to the
associative ionization. Instead, it was proportional to Na$^{+}$ signal, and should be therefore considered as
some kind of technical noise originating from Na$^{+}$ ions.

Analysis of the experimental arrangement used in the present study has shown that the noise was related to the
Na$^{+}$ ions produced during the second extraction electric field pulse (figure~3(b)). The pulse acted on these
ions for a shorter time than on Na$^{+}$ ions produced before the electric pulse. Therefore, the former ions
reached the channeltron at a later time than the latter ones. As a result, the detection gate, which was set for
registration of the molecular ions, counted also the Na$^{+}$ ions produced during the second extraction field
pulse. This noise could not be eliminated by any technical means, and unavoidably had to be subtracted from the
total ion signal counted during the molecular ion registration gate, although it was never reported in earlier
experiments. The magnitude of this noise signal can be expressed by the formula

\begin{figure}
\begin{center}
\epsfxsize=13 cm \epsfbox{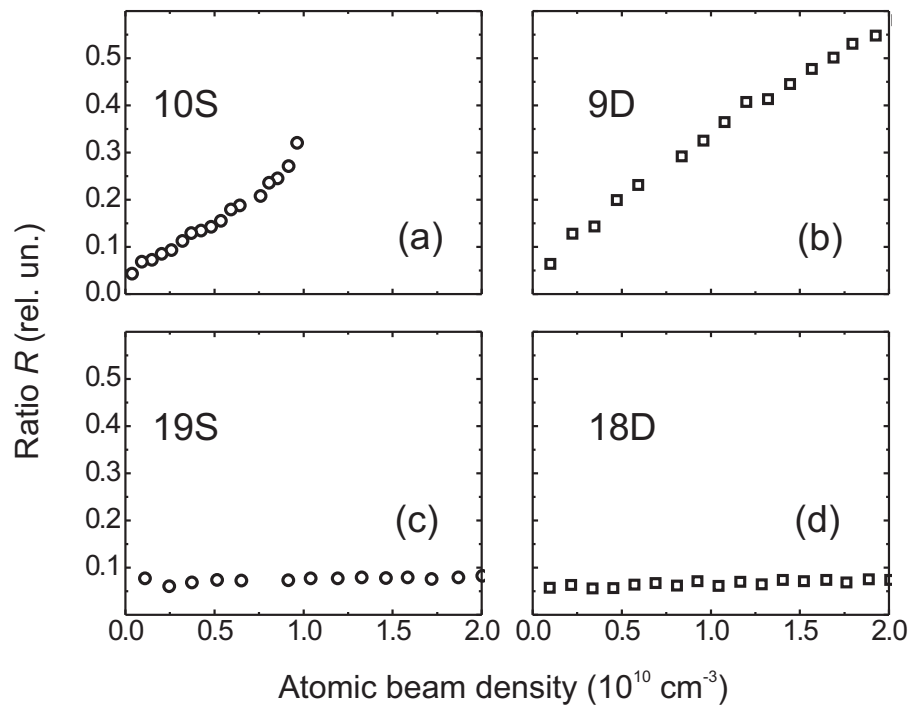} \caption{{\label{Fig8}Experimental dependences of the ratio $R$ of Na$_2^+$
and Na$^+$ ion signals on the density $n_{3S}$ of ground state atoms in the beam for the (a) 10S state, (b) 9D
state, (c) 19S state, and (d) 18D state.}}
\end{center}
\end{figure}
\begin{equation}
\label{eq20}
N_{noise} = N_{nL} \left( {0} \right)\;W_{BBR}^{tot} \;e^{ - t_{2} /\tau
_{eff}} \Delta t,
\end{equation}

\noindent where $\Delta $\textit{t} is the time interval of the second electric extraction pulse, during which
those Na$^{+}$ ions are created, which can fall into the gate for molecular ions, and the time \textit{t}$_{2}$
corresponds to the beginning of the extraction pulse. The noise is proportional to $W_{BBR}^{tot} $, which grows
rapidly with increasing \textit{n} (figure~6). The time interval $\Delta $\textit{t} was calculated from the
geometry of the detection system and the duration of the gate counting the molecular ions, and it was found to be
about 150 ns. To find the contribution $\alpha $ of the noise to the value of \textit{R}, equation~(\ref{eq20})
must be divided by equation~(\ref{eq4}) for Na$^{+}$ ions:

\begin{equation}
\label{eq21}
\alpha = \frac{{\Delta t}}{{\tau _{eff}} }\frac{{1}}{{e^{t_{2} /\tau _{eff}
} - 1}}.
\end{equation}

When this contribution is known, the absolute rate constant for AI in Na(\textit{nL})+Na(3S) collisions can be
extracted from the experimental ratios \textit{R} of the molecular and atomic ion signals using
equation~(\ref{eq10}) modified by the correction $\alpha $ of equation~(\ref{eq21}):

\begin{equation}
\label{eq22}
k_{AI} = \left( {R - \alpha}  \right) \cdot \frac{{W_{BBR}^{tot}} }{{n_{3S}
}}.
\end{equation}

The values of $\alpha $ for \textit{n}S and \textit{n}D states were calculated using equation~(\ref{eq21}). For
highest \textit{n} of the experiment they practically coincide with the measured ratios (figure~8(c) and 8(d))

\subsection{AI rate constants}

 The experimental dependences of \textit{k}$_{AI}$ on \textit{n}$_{eff}$ were obtained from equation~(\ref{eq22})
using the measured ratios of Na$_{2}^{+}$ and Na$^{+}$ signals and the data of Table 1. These dependences are
shown in figure~9. For the sake of completeness, we present two sets of results. Figure 9(a) shows the rate
constants determined with the experimental values $W_{BBR}^{tot} $ of the BBR ionization rate, while figure~9(b)
shows the same but using the theoretical $W_{BBR}^{exp} $ values. The plots in both figures are nearly identical,
except for some minor differences in the absolute\textit{} values of \textit{k}$_{AI}$. Noteworthy, all
dependences show the maximum in the AI rate near $n_{eff}\approx (11-12)$, and fall off rapidly with
increasing\textit{ n}$_{eff}$.

\begin{figure}
\begin{center}
\epsfxsize=15 cm \epsfbox{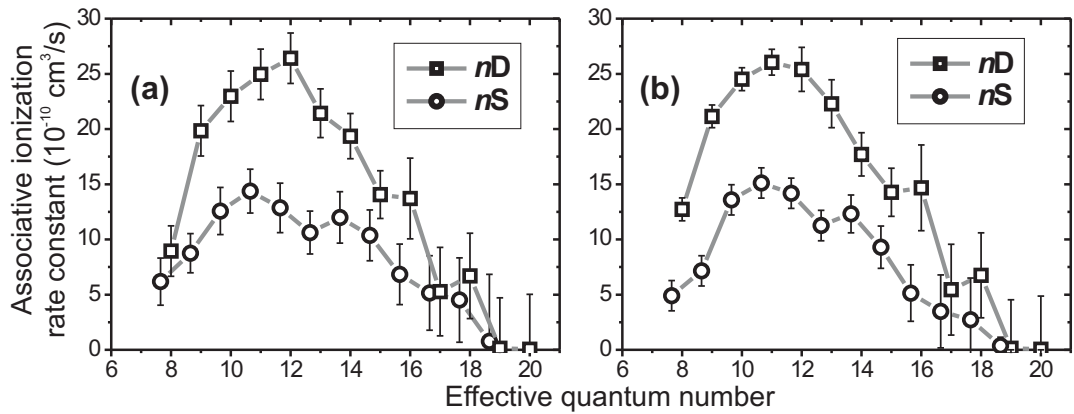} \caption{{\label{Fig9}Experimental data on the associative ionization rates
for \textit{n}S and \textit{n}D states, obtained from equation~(\ref{eq22}) using the measured ratio of
Na$_{2}^{+} $ and Na$^{+}$ ion signals, and (a) the experimental values of $W^{tot}_{BBR}$ ; (b) the theoretical
values of $W^{tot}_{BBR}$}}
\end{center}
\end{figure}

Comparing our results with those of earlier AI studies, shown in figure~1, we see a noticeable difference with the
earlier data of \cite{Weiner1986} on \textit{k}$_{AI}$ for \textit{n}S and \textit{n}D states which were also
obtained in a single beam experiment: (i) in our study the rate constants of AI for \textit{n}D states are nearly
twice larger than those for \textit{n}S states, while opposite ratio was reported in \cite{Weiner1986}, and (ii)
our values of \textit{k}$_{AI}$ fall off more rapidly with increasing \textit{n} than those of \cite{Weiner1986}.
We believe that our measurements are more reliable, since, in contrast to \cite{Weiner1986}, using
equation~(\ref{eq22}) we did not need to measure the number of Rydberg atoms and the effective lifetimes, and thus
we could avoid the related experimental uncertainties.

The only important parameter, which was not exactly known in the present experiment, was the density
\textit{n}$_{3S}$ of ground state atoms in the beam, which was calculated using the vapor pressure formula of
Nesmeyanov \cite{Nesmeyanov}. At the same time, care was taken to ensure that the source temperature and the
density of atoms in the beam were constant during the experiment. Therefore, the relative values of
\textit{k}$_{AI}$ for different \textit{n}S and \textit{n}D states are not affected by this uncertainty.

Another interesting observation can be made by comparing our data of figure~9(b) with the \textit{n} dependence
for \textit{n}P states of figure~1. Such comparison is given in figure~10. One can see that both data sets have
similar absolute values and similar dependences on the effective quantum number \textit{n}$_{eff}$\textit{.} Note,
however, that the data reported for \textit{n}P states in \cite{Weiner1986} consist actually of the AI rate
constants for the states with 5$ \le $\textit{n$ \le $}15, which were measured in \cite{Boulmer1983} under quite
different conditions of a crossed beam experiment, and the rate constants for $n>15$, which were actually measured
in a single beam experiment \cite{Weiner1986}. Joining of the two data sets was used in \cite{Weiner1986} for
scaling their relative cross section values for $n > 15$.

\begin{figure}
\begin{center}
\epsfxsize=9 cm \epsfbox{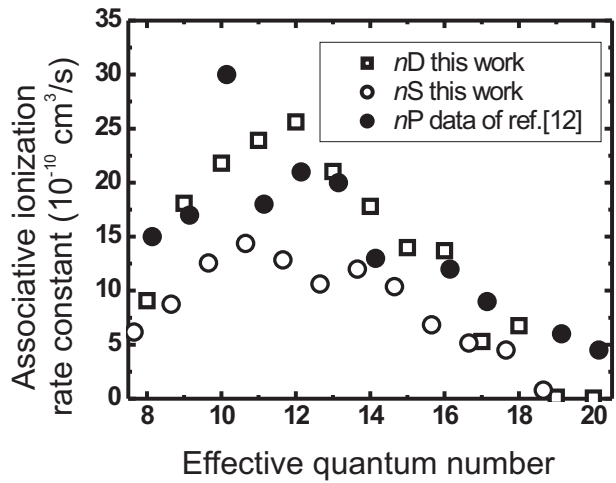} \caption{{\label{Fig10}Comparison of the experimental associative ionization
rate constants for $n$S (open circles) and $n$D (open squares) states obtained in the present study (figure~9(b))
with the experimental rate constants for $n$P states (solid circles) obtained in \cite{Boulmer1983} and
\cite{Weiner1986} (figure~1). }}
\end{center}
\end{figure}

\section{Discussion}

Our measurements of associative ionization rate constants were based on use of equation~(\ref{eq22}). The accuracy
of the measured \textit{k}$_{AI}$ values depends on the accuracy with which the values of parameters \textit{R},
$\alpha $, $W_{BBR}^{tot} $, and \textit{n}$_{3S}$ are known. The ratio \textit{R} of Na$_{2}^{+}$ and Na$^{+}$
ions was measured directly and was found to be independent of laser intensities, detunings, and effective
lifetimes. The measured \textit{n} dependences of figure~7 were reproduced in several experiments, which allow us
to be confident that the values of \textit{R} are determined correctly. However, with increasing \textit{n} the
fluctuations of \textit{R} became important, because equation~(\ref{eq22}) uses a product of (\textit{R}-$\alpha
$), which decreases for high \textit{n} to values close to 0, and $W_{BBR}^{tot} $, which rapidly grows with
increasing \textit{n}. Therefore the reported values of \textit{k}$_{AI}$ can be considered as reliable for lower
\textit{n$ \le $}16, whereas for higher \textit{n} the error of determination of \textit{k}$_{AI}$ is relatively
large. More reliable data for higher \textit{n} could be obtained by increasing the density of atoms in the beam
by an order of magnitude or more. Unfortunately such increase inevitably leads to complications in the analysis
due to additional phenomena, like radiation trapping, which was found to strongly affect the velocity
distributions of excited atoms in a beam experiment at densities of 2$ \times $10$^{11}$~cm$^{-3}$
\cite{Kauffman}.

Another possible source of errors in the determination of \textit{k}$_{AI}$ is the choice of values of the total
BBR induced ionization rate $W_{BBR}^{tot} $. Figure~6 shows a good agreement of experimental and theoretical
values for all \textit{n}D states, while for higher \textit{n}S states the agreement is rather poor. Therefore,
one must choose whether the theoretical or the experimental values of BBR ionization rates, given by the last two
columns of Table 1, will be used in equation~(\ref{eq22}). Since the experimental results of figure~6 were
reproduced in many experiments, we believe that the experimental values $W_{BBR}^{exp} $ are more reliable.
Therefore, figure~9(a) should be considered as the final result of the \textit{k}$_{AI}$ measurements.

The absolute values of \textit{k}$_{AI}$ depend also on the absolute value of \textit{n}$_{3S}$ calculated using
the Nesmeyanov's formula \cite{Nesmeyanov}. Numerous studies involving alkali atoms have confirmed its accuracy in
predicting the vapor pressures of alkali gases (see, e.g., the argumentation in \cite{Ekers}) within a few tens of
per cent. Note, that the relative values of \textit{k}$_{AI}$ were not affected by the absolute value of
\textit{n}$_{3S}$, since it was kept constant during the experiment. We conclude that on average our absolute
values of \textit{k}$_{AI}$ are accurate to within better than 50\%, while the relative values of
\textit{k}$_{AI}$ are accurate to within (10-20)\%. However, due to the specific \textit{n} dependence of the term
$(R-\alpha )$ in equation~(\ref{eq22}), the reliability of \textit{k}$_{AI}$ values is better for lower
\textit{n}.

In conclusion, we have measured, for the first time, AI rate constants for sodium \textit{n}S and \textit{n}D
states in the range of 8$ \le $\textit{n$ \le $}20. In this study, we have implemented a novel method, in which
the absolute values of rate constants were determined from the measurements of relative ratios of the molecular
and atomic ion signals, with atomic ions being from the BBR ionization of Rydberg atoms. The main advantage of
this approach is that the measurements of absolute populations and effective lifetimes, which are associated with
large experimental uncertainties, can be avoided.

To implement this method, we have measured BBR ionization rates for \textit{n}S and \textit{n}D states of sodium
with \textit{n} = 8-20. Comparison of these data with the theory showed a good agreement for \textit{n}D states,
and disagreement for \textit{n}S states with $n>15$. This discrepancy cannot be attributed to any of the
experimental uncertainties, and the theory must substantially revise the calculations for \textit{n}S states.

The comparison of our experimental data on AI rates with other experiments for \textit{n}S and \textit{n}D states
is hampered by lacking overlap of the studied ranges of \textit{n}. However, for the high \textit{n} range, where
some overlap with the data of [12] exists, notable differences were observed. This may be explained by the way we
accounted for the background Na$^{+}$ ions. These ions were created due to ionization during the extraction
electric field pulse, and they arrived to the detector simultaneously with Na$_{2}^{+}$. Such possibility was not
considered in earlier studies \cite{Boulmer1983,ZagrebinSamsonZ,ZagrebinSamsonJ,Weiner1986}, and it is not
possible to reconstruct whether such background was present in these experiments.

Curiously, our results for \textit{n}S and \textit{n}D states are in a good agreement with the data of
\cite{Boulmer1983} for \textit{n}P states, though that experiment was performed with two crossed atomic beams. In
the present single-beam study, ionization is due to head-tail overtaking collisions between ground-state and
Rydberg atoms. Accordingly, the collision energies are about three times lower than those in the crossed-beam
experiment, which used beam sources of the same temperature. In the second part of our article \cite{Miculis} we
provide a detailed comparison of the existing experimental data on associative ionization rate constants with the
theoretical calculations for various relative velocity distributions. The calculations are based on the stochastic
ionization model \cite{Bezuglov1999,Bezuglov1997,BezuglovEkers2002,Bezuglov2003,Bezuglov2002}, and they take into
account all possible collisional, spontaneous, and BBR-induced mixing processes.

\section{Acknowledgments}

This work was supported by INTAS Project No.2001-155, the Russian Foundation
for Basic Research, Grant No.02-02-16332, and the Deutsche
Forschungsgemeinschat.


\section{References}
\begin{thebibliography}{20}
\bibitem{Janev1980}Janev~R~K and Mihajlov~A~A 1980 \textit{Phys.~Rev.~A} \textbf{21} 819
\bibitem{Mihajlov1981}Mihailov~A~A and Janev~R~K 1981 \textit{J.~Phys.~B} \textbf{14} 1639
\bibitem{DumanShmatov}Duman~E~L and Shmatov~I~P 1980 \textit{Sov. Phys. JETP} \textbf{51} 1061
\bibitem{Bezuglov1999}Bezuglov~N~N, Borodin~V~M, Klyucharev~A~N, Fuso~F, Allegrini~M,
Orlovsky~K~V and Janson~M~L 1999 \textit{Opt.~Spektrosc.} \textbf{86} 824
\bibitem{BezuglovEkers2002}Bezuglov~N~N, Borodin~V~M, Ekers~A and Klyucharev~A~N 2002
 \textit{Opt.~Specrosc.} \textbf{93} 661
\bibitem{Bezuglov2003}Bezuglov~N~N, Borodin~V~M, Grushevskyi~V, Klyucharev~A~N,
Michulis~K, Fuso~F and Allegrini~M 2003 \textit{Opt.~Spectrosc.} \textbf{95} 515
\bibitem{Bezuglov2002}Bezuglov~N~N, Borodin~V~M, Kazanskii~A~K, Klyucharev~A~N, Matveev~A~A and
Orlovskii~K~V 2002 \textit{Opt. Spectrosc.} \textbf{91} 19
\bibitem{Bezuglov1997}Bezuglov~N~N, Borodin~V~M, Klyucharev~A~N, Orlovsky~K~V and
Allegrini~M 1997 \textit{Opt.~Spectrosc.} \textbf{82} 334
\bibitem{Boulmer1983}Boulmer~J, Bonanno~R and Weiner~J 1983, \textit{J.~Phys.~B} \textbf{16} 3015
\bibitem{ZagrebinSamsonZ}Zagrebin~S~B and Samson~A~V 1984 \textit{Pisma Zh. Tekhn. Fiz.} \textbf{10} 114
(in Russian)
\bibitem{ZagrebinSamsonJ}Zagrebin~S~B and Samson~A~V 1985 \textit{J.~Phys.~B} \textbf{18} L217
\bibitem{Weiner1986}Weiner~J and Boulmer~J 1986 \textit{J.~Phys.~B} \textbf{19} 599
\bibitem{Burkhardt1984}Burkhardt~C~E, Garver~W~P, Kushawaha~V~S and
Leventhal~J~J 1984 \textit{Phys. Rev. A} \textbf{30} 652
\bibitem{Burkhardt1986}Burkhardt~C~E, Corey~R~L, Garver~W~P, Leventhal~J~J,
Allegrini~M and Moi~L 1986 \textit{Phys.~Rev.~A} \textbf{34} 80
\bibitem{Klucharev}Klucharev~A~N and Janson~M~L \textit{Elementary processes in the
plasma of alkali metals} (Energoatomizdat, Moscow, 1988) (in Russian).
\bibitem{Theodosiou}Theodosiou~C~E 1984 \textit{Phys.~Rev.~A} \textbf{30} 2881
\bibitem{Gallagher}Gallagher~T~F \textit{Rydberg atoms} (Cambridge University
Press, Cambridge, 1994)
\bibitem{Spencer}Spencer~W~P, Vaidyanathan~A~G, Kleppner~D and Ducas~T~W, 1982 \textit{
Phys.~Rev.~A} \textbf{26}, 1490
\bibitem{Lehman}Lehman~G~W 1983 \textit{Phys.~Rev.~A} \textbf{16} 2145
\bibitem{Allegrini1988}Allegrini~M, Arimondo~E, Menchi~E, Burkhardt~C~E, Ciocca~M,
Garver~W~P, Gozzini~S, Leventhal~J~J and Kelley~J~D 1988 \textit{Phys.~Rev.~A} \textbf{38} 3271
\bibitem{Cooke}Cooke W~E and Gallagher~T~F 1980 \textit{Phys. Rev. A}
\textbf{21} 588
\bibitem{Galvez}Galvez~E~J, Lewis~J~R, Chaudhuri~B, Rasweiler~J~J, Latvakoski~H, De~Zela~F, Massoni~E and
Castillo~H 1995 \textit{Phys. Rev. A} \textbf{51} 4010
\bibitem{Ryabtsev}Ryabtsev~I~I, Tretyakov~D~B, Beterov~I~I 2003 \textit{J.~Phys.~B} \textbf{36} 297
\bibitem{Nesmeyanov}Nesmeyanov~A~N \textit{Vapur pressure of the chemical elements}
(Elsevier, Amsterdam / London / New York, 1963)
\bibitem{PhysicalValues}Handbook \textit{Physical values} /Edited by Grigoryev~I (Energoatomizdat, Moscow, 1991)(in Russian).
\bibitem{BoulmerDimer} Boulmer~J and Weiner~J, 1983 \textit{Phys.~Rev.~A} \textbf{27} 2817
\bibitem{Burkhardt1985}Burkhardt~C~E, Garver~W~P and Leventhal~J~J, 1985 \textit{Phys.~Rev.~A} \textbf{31} 505
\bibitem{Meijer}Meijer~H~A~J, Pelgrim~T~J~C, Heideman~H~G~M, Morgensten~R and Andersen~N 1989 \textit{J. Chem. Phys.} \textbf{90} 738;
Meijer~H~A~J 1990 \textit{Z. Phys. D} \textbf{17} 257; Meijer~H~A~J, van der Meulen~H~P and Morgenster~R 1987
\textit{ibid} \textbf{5} 299; Meijer~H~A, Schoh~S, Muller~M~W, Dengel~H,  Ruf~M~-W and Hotop~H 1991 \textit{J. Phys. B}
\textbf{24} 3621

\bibitem{Dulieu}Huynh~B, Dulieu~O, and Masnou-Seeuws~F 1998 \textit{Phys. Rev. A} \textbf{57} 958
\bibitem{Blange}Blange~J~J, Urbain~X, Rudolph~H, Dijkerman~H~A, Beijerinck~H~C~W, and Heideman~H~G~M 1997 \textit{J. Phys. B} \textbf{30} 565
\bibitem{SFI} Stebbings~R~F, Latimer~C~J, West~W~P, Dunning~F~B and Cook~T~B
1975 \textit{Phys.~Rev.~A} \textbf{12} 1453; Ducas~T, Littman~M~G, Freeman~R~R and Kleppner~D 1975
\textit{Phys.~Rev.~Lett.} \textbf{35} 366; Gallagher~T~F, Humphrey~L~M, Hill~R~M and Edelstein~S~A 1976
\textit{Phys.~Rev.~Lett.} \textbf{37} 1465
\bibitem{DyachkovPankratov}Dyachkov~L~G and Pankratov~P~M 1994 \textit{J.~Phys.~B} \textbf{27} 461
\bibitem{Aymar}Aymar~M 1978 \textit{J.~Phys.~B} \textbf{11} 1413
\bibitem{Zimmerman}Zimmerman~M~L, Littman~M~G, Kash~M~M and Kleppner~D 1979
\textit{Phys.~Rev.~A} \textbf{20} 2251
\bibitem{BezuglovBorodin}Bezuglov~N~N and Borodin~V~M 1999 \textit{Opt.~Spectrosc.} \textbf{86} 467
\bibitem{Dyubko}Dyubko~S~F, Efimenko~M~N, Efremov~V~A, Podnos~S~A 1995
\textit{Quantum Electronics} \textbf{25} 914
\bibitem{Kauffman}Kaufmann~O, Ekers~A, Bergmann~K, Bezuglov~N, Miculis~K, Auzinsh~M
and Meyer~W 2003 \textit{J.~Chem.~Phys} \textbf{119} 3174.
\bibitem{Ekers}Ekers~A, Glodz~M, Szonert~J, Bieniak~B, Fronc~K and
Radelitski~T 2000 \textit{Eur.~Phys.~J.~D} \textbf{8} 49
\bibitem{Miculis}Miculis~K, Beterov~I~I, Bezuglov~N~N, Ryabtsev~I~I, Tretyakov~D~B,
Ekers~A, Klyucharev~A~N 2005 \textit{J.~Phys.~B} (submitted).

\end{thebibliography}
\end{document}